\documentclass[aps,physrev,preprint, superscriptaddress]{revtex4-2}

\usepackage{graphicx}
\usepackage{booktabs} 
\usepackage{tabularx} 
\usepackage{xcolor}
\usepackage{siunitx}

\newcommand{\bochum}{Research Center Future Energy Materials and Systems of the University Alliance Ruhr and Interdisciplinary Centre for Advanced Materials Simulation, Faculty of Physics and Astronomy, Ruhr University Bochum, Universitätsstraße 150, D-44801 Bochum, Germany}
\newcommand{\jena}{Institut f\"ur Festk\"orpertheorie und -optik, Friedrich-Schiller-Universit\"at Jena, Max-Wien-Platz 1, 07743 Jena, Germany}
\newcommand{\etsf}{European Theoretical Spectroscopy Facility (ETSF)}

\begin{document}


\title{Optical selection rules in hexagonal Ge polytypes and their lifting by symmetry perturbation}


\author{Martin Keller}
 \affiliation{\jena}
 \affiliation{\etsf}
\author{Haichen Wang}%
 \affiliation{\bochum}
 \affiliation{\etsf}
\author{Friedhelm Bechstedt}%
 \affiliation{\jena}
 \affiliation{\etsf}

\author{J\"urgen Furthmüller}
 \affiliation{\jena}
\affiliation{\etsf}

\author{Silvana Botti}
\affiliation{\bochum}
\affiliation{\jena}
\affiliation{\etsf}

\date{\today}

\begin{abstract}
Hexagonal germanium polytypes have emerged as promising direct-gap semiconductors for silicon-integrated optoelectronics, yet their optical properties remain largely unexplored beyond the well-studied 2H phase. We present a comprehensive theoretical study of optical properties of hexagonal 2H-, 4H-, and 6H-Ge polytypes through ab initio calculations of quasiparticle band structures, dipole transition matrix elements, and solution of the Bethe-Salpeter equation. While all three polytypes exhibit direct band gaps of increasing size from 2H to 6H, we reveal that the fundamental optical transition in 4H-Ge is parity-forbidden due to matching band parities at the valence and conduction band edges. This selection rule results in a radiative lifetime seven orders of magnitude longer than in 2H- and 6H-Ge, severely limiting light emission capabilities. To demonstrate that the selection rule can be lifted, we introduce controlled symmetry perturbations by substituting single Ge atoms with Si in each unit cell, breaking the crystal symmetry. This perturbation increases the optical matrix elements by up to two orders of magnitude and reduces radiative lifetimes for all perturbed polytypes. We also compute absorption coefficients and frequency-dependent dielectric tensors for both light polarizations, including excitonic effects up to 5 eV, providing complete optical characterization of ideal and symmetry-perturbed hexagonal Ge systems relevant for optoelectronic applications.
\end{abstract}

\maketitle


\section{Introduction}

Silicon-based microelectronics and group III-V communication technologies remain fundamentally incompatible. Silicon is abundant and inexpensive, but its indirect band gap prevents efficient light emission. Meanwhile, group III-V materials resist integration with existing Si CMOS platforms. Bridging this divide has been a central challenge in photonics for decades \cite{Shen2019,shekhar2024roadmapping}, with strategies ranging from III-V integration to Ge strain engineering, each facing significant obstacles.

Recent experimental breakthroughs have demonstrated that hexagonal germanium offers a promising solution. Hexagonal Ge with lonsdaleite crystal structure (2H-Ge) and Ge-rich hexagonal SiGe alloys exhibit direct band gaps with lattice structures compatible with Si \cite{nature,nanosecondLifetimeHexGe}. These materials achieve light emission comparable to group III-V semiconductors while maintaining compatibility with silicon technology. The tunability of the emission wavelength through alloy composition, ranging from \SI{1.8}{\micro\meter} to \SI{3.4}{\micro\meter}, makes them particularly attractive for integrated optoelectronic applications. Alloying with Si below $\SI{45}{\%}$ \cite{nature} and 2H-Ge quantum wells \cite{Peeters2024Direct} both demonstrate strong optical transitions in photoluminescence (PL) spectra at the band gap.

However, the optical activity of hexagonal group-IV materials is not straightforward. Pure 2H-Ge is a pseudo-direct semiconductor: while it possesses a direct fundamental band gap, the optical transitions at this minimum energy are extremely weak due to symmetry constraints \cite{2HClaudia,alloyTransitionsPedro}. This weak optical activity manifests in calculated radiative lifetimes exceeding \SI{0.1}{ms} at low temperatures \cite{2HClaudia}, several orders of magnitude longer than typical direct-gap semiconductors. The pseudo-direct nature arises from dipole selection rules that suppress the oscillator strength of the fundamental transition, severely limiting the material's potential for light-emitting devices. Breaking the crystal symmetry through alloying or strain has been shown to lift these selection rules dramatically, enhancing oscillator strengths by orders of magnitude \cite{alloyTransitionsPedro,2Helements,PhysRevMaterials.5.024602}. Even small Si incorporation triggers symmetry reduction, enabling efficient radiative recombination with nanosecond lifetimes \cite{nature}.

The optical properties of hexagonal materials are accessible through experimental characterization and theoretical calculations. Recent electron energy loss spectroscopy (EELS) studies of 2H-Ge and 2H-Si \cite{absorptionEELS} reveal strong optical absorption above \SI{2}{eV} (2H-Ge) or \SI{3}{eV} (2H-Si), though the low-energy region below \SI{1}{eV} remains inaccessible to this technique. Complete optical spectra, i.e., the real and imaginary parts of the frequency-dependent dielectric tensor for both light polarizations, have been calculated for 2H-Si and 2H-Ge including many-body effects \cite{absorptionEELS,PhysRevB.92.045207,alloyVibrationTheory}, as well as for 2H-SiGe alloys at the independent-particle level \cite{PhysRevMaterials.7.014602}.

Beyond the well-studied 2H phase, hexagonal polytypes 4H- and 6H-Ge have been predicted theoretically to possess direct band gaps \cite{polytypes}. Experimental progress includes the growth of 4H-Ge nanowires \cite{thesis:victor} and Raman identification of the 6H polytype in Ge films on AlGaAs nanowires \cite{6Hraman}. Synthesis routes for 4H polytypes have been demonstrated through solution-phase reactions for Ge \cite{B921575A} and phase transformations under heating or pressure for Si \cite{PhysRevLett.126.215701,doi:10.1021/acs.nanolett.8b02816}. While approximate quasiparticle calculations have characterized the electronic properties of 4H- and 6H-Ge \cite{polytypes}, their optical properties remain largely unexplored. If these polytypes exhibit light emission behavior comparable to 2H-Ge, they would significantly expand the material space available for heterostructure design. However, preliminary data reveal a puzzling discrepancy: the radiative lifetime of 4H-Ge reaches $\tau = \SI{0.121}{ms}$ \cite{thesis:victor}, seven orders of magnitude longer than the $\tau = \SI{6}{ns}$ measured in 2H-Ge at low temperatures \cite{nature}. This dramatic variation demands explanation.

In this work, we provide a comprehensive investigation of optical properties in hexagonal Ge polytypes, focusing on both the infrared region near the absorption edge and the broader ultraviolet range. We first compute and analyze optical transition strengths for 4H- and 6H-Ge using \textit{ab initio} electronic-structure methods, comparing them to the 2H polytype. Our calculations reveal that the lowest energy transition of 4H-Ge is parity-forbidden, explaining its extremely long radiative lifetime. We then demonstrate how symmetry perturbations can lift this selection rule by studying Ge polytypes with single Si substitutions per unit cell. We quantify the light emission capabilities through radiative lifetime calculations and characterize absorption edges using many-body perturbation theory. In the second part, we extend our analysis to higher photon energies, computing frequency-dependent dielectric functions for both light polarizations with full treatment of quasiparticle and excitonic effects.

\section{Theoretical and Numerical Approaches}

We performed all calculations using the \textit{Vienna ab initio Simulation Package} (VASP) \cite{vasp1,vasp2}, employing the projector-augmented wave (PAW) method \cite{PAW} to describe wavefunctions and pseudopotentials. We set the plane wave expansion cutoff at \SI{500}{eV} and performed Brillouin-zone (BZ) integrations on $\Gamma$-centered grids of $12\times 12\times 6$, $12\times 12\times 3$, and $12\times 12\times 2$ for the 2H, 4H, and 6H structures, respectively. This choice ensures uniform grid density across all polytypes, yielding total energy errors below 1\,meV.

For structural optimization of the SiGe alloys, we applied the PBEsol variant of the Perdew-Burke-Ernzerhof (PBE) exchange-correlation functional \cite{PBEsol}, relaxing atomic positions until Hellmann-Feynman forces fell below \SI{0.1}{meV/\AA}. We then computed electronic band structures and transition matrix elements using the MBJLDA meta-GGA functional \cite{BeckeJohnson,TranBlaha1,TranBlaha2} with spin-orbit coupling (SOC) included. This meta-GGA approach provides approximate quasiparticle (QP) band structures \cite{manybodyElectronicExcitations,borlido2019large}, significantly improving upon Kohn-Sham band structures obtained with PBEsol.

To characterize optical transitions between conduction band $c_m$ and valence band $v_n$, we computed dipole matrix elements averaged over spin-orbit degenerate states $i,j = 1,2$ \cite{manybodyElectronicExcitations,2HClaudia}:
\begin{equation}
    p^{\perp/\parallel}_{m,n}(\mathbf{k}) = \sqrt{\frac{1}{2}\sum_{i,j}{\left|\left<c_{mi},\mathbf{k}\left|p^{\perp/\parallel}\right|v_{nj},\mathbf{k}\right>\right|^2}}\text{,}
    \label{eq:momentum_matrix}
\end{equation}
where $\perp/\parallel$ denotes light polarization perpendicular or parallel to the hexagonal c-axis, $p$ represents the momentum operator, and $\left|c/v\mathbf{k}\right>$ are the Bloch functions. For SiGe alloys, we further averaged matrix elements over multiple unique unit cell configurations and, for the perpendicular component $p_\perp$, over the $x$ and $y$ Cartesian directions.

We characterized the optical properties through the two diagonal elements of the dielectric tensor $\epsilon_{\perp/\parallel}(\omega)$, computed by solving the Bethe-Salpeter equation (BSE) \cite{manybodyElectronicExcitations} to include excitonic effects. Accurate description of the imaginary parts of the dielectric function up to \SI{5}{eV} required including transitions over 192, 384, and 576 bands for the 2H, 4H, and 6H polytypes, respectively. We used the same $\Gamma$-centered k-point grids for the optical spectra: $12\times 12\times 6$, $12\times 12\times 3$, and $12\times 12\times 2$ for the 2H, 4H, and 6H structures, respectively. From the calculated real and imaginary parts of $\epsilon_{\perp/\parallel}(\omega)$, we derived the absorption coefficients $\alpha_{\perp/\parallel}(\omega)$ for both light polarizations.

\begin{figure}
    \centering
    \includegraphics[width=0.6\textwidth]{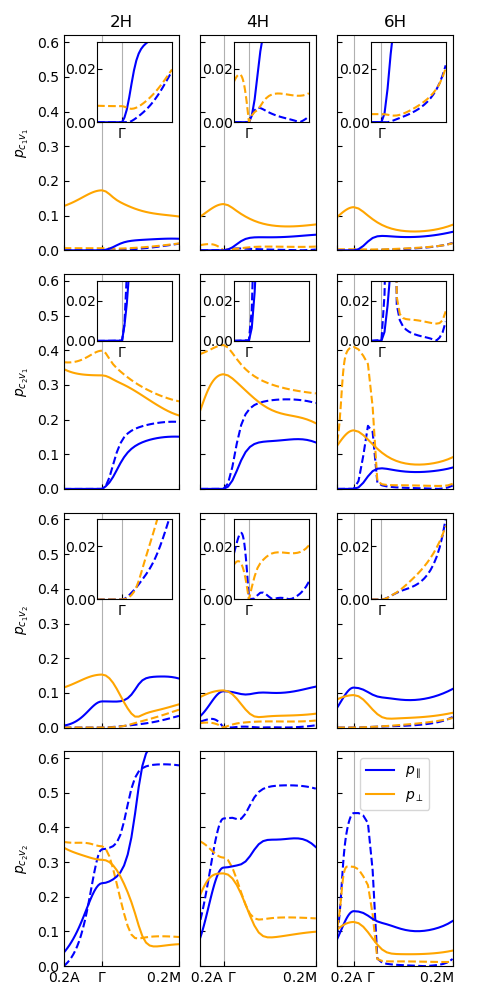}
    \caption{Optical transition matrix elements of pure (dashed lines) and alloyed (solid lines) Ge polytypes in units of $\hbar /a_B$ near the $\Gamma$-point, shown for light polarization perpendicular and parallel to the hexagonal c-axis. Insets provide magnified views of the weaker transitions in pure Ge.}
    \label{fig:matrix-elements}
\end{figure}

\section{Optical Transitions near Band Edge}

\subsection{Matrix Elements and Selection Rules}

Figure \ref{fig:matrix-elements} displays the optical transition matrix elements for the three Ge polytypes (dashed lines), where $c_i$ ($v_i$) denotes the $i$th-lowest conduction (highest valence) band. The behavior of 2H-Ge matrix elements is well established \cite{2HClaudia,2Helements,kpHexGe}: the lowest conduction band exhibits weak optical activity, with transition strengths between $c_1$ and the highest valence bands being two orders of magnitude smaller than those involving $c_2$. These transitions are forbidden by dipole selection rules (cf. Table \ref{tab:matrix_elements} and Fig. \ref{fig:matrix-elements}), though spin-orbit coupling, the hexagonal crystal field, and alloying can soften these restrictions, enabling weak transitions \cite{kpHexGe}. 

The behavior and magnitude of matrix elements near the $\Gamma$-point are remarkably similar across all three polytypes, including the characteristic weak optical activity of $c_1$. Since all pure Ge polytypes share the point group $\mathrm{D}_{6\mathrm{h}}$, the same selection rules apply for band states of identical symmetry \cite{2HClaudia}. However, 4H-Ge presents a striking exception: transitions involving its lowest conduction band $c_1$, particularly the fundamental transition, are strictly forbidden at $\Gamma$, even for light polarized perpendicular to the c-axis. This distinguishes 4H-Ge from the 2H and 6H systems, despite sharing similar matrix element magnitudes away from $\Gamma$. The gap energies $E_{11}$ (corresponding to the transition $v_1 \rightarrow  c_1$) listed in Table \ref{tab:matrix_elements} increase systematically along the series 2H$\rightarrow$4H$\rightarrow$6H. Our calculated 2H value of \SI{0.3}{eV} agrees well with other approximate quasiparticle calculations employing hybrid exchange-correlation functionals \cite{2HClaudia,absorptionEELS,alloyVibrationTheory,doi:10.1021/acs.jpcc.6b12782,Tizei2020,Broderick2024arxiv,zq7r-s182}.

\begin{figure}
    \centering
    \includegraphics[width=0.5\textwidth]{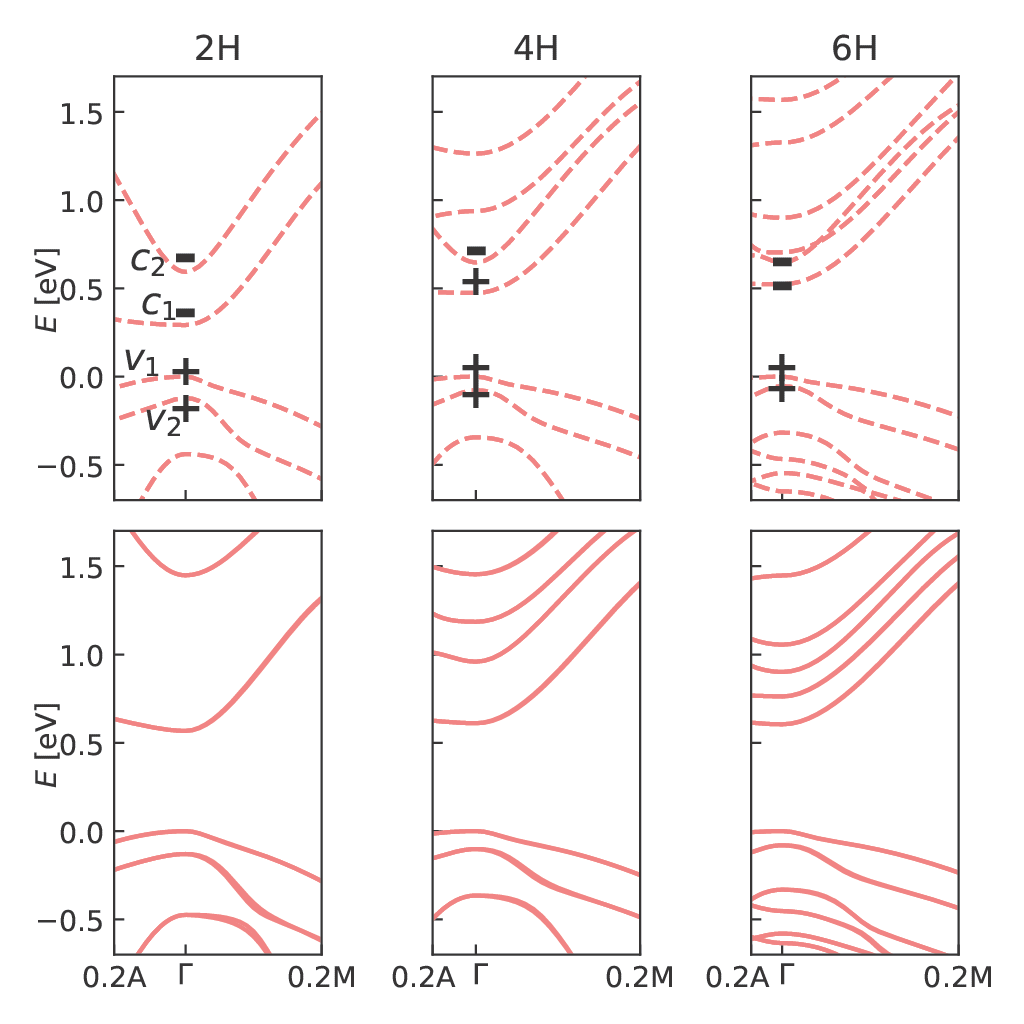}
    \caption{Electronic band structures of pure (dashed lines) and alloyed (solid lines) Ge polytypes near the $\Gamma$-point. Signs indicate band parities. The valence band maximum defines the energy zero.}
    \label{fig:bandstructures}
\end{figure}

To identify the selection rule responsible for this forbidden transition, we analyzed the electronic structure symmetries by computing parity operator eigenvalues at the $\Gamma$-point using the IrRep code \cite{IrRep}. Optical transitions are allowed between states of opposite parity and forbidden between states of the same parity. Figure \ref{fig:bandstructures} shows the band structures near $\Gamma$, with band parities indicated by signs. For all three polytypes, the two highest valence bands possess parity $+1$. The $c_1$ band of 4H-Ge also has parity $+1$, while the corresponding bands in 2H- and 6H-Ge have parity $-1$. Consequently, transitions involving $c_1$ in 4H-Ge are parity-forbidden. This forbidden fundamental transition disqualifies pure 4H-Ge as a candidate for active optoelectronic applications.

Since these forbidden transitions arise from symmetry properties, appropriate perturbations can lift them by breaking inversion symmetry. We model this through SiGe alloying \cite{2Helements,alloyTransitionsPedro}, replacing one Ge atom per unit cell with Si. For the 4H (6H) structures, there are 2 (3) unique substitution sites. We computed matrix elements for each configuration and averaged the results, yielding alloy compositions 2H-Si$_{0.25}$Ge$_{0.75}$, 4H-Si$_{0.125}$Ge$_{0.875}$, and 6H-Si$_{0.083}$Ge$_{0.917}$. 

Figure \ref{fig:matrix-elements} shows the alloy matrix elements (solid lines). Breaking the inversion symmetry reduces the point group to $\mathrm{C}_{3\mathrm{v}}$, removing the parity-forbidden character of 4H transitions involving $c_1$. Moreover, transitions involving $c_1$ in all polytypes now reach the same order of magnitude as other transitions in both alloys and pure Ge. However, alloy matrix elements involving $c_2$ decrease by 11--64\% compared to pure hexagonal Ge, with the largest reduction occurring in the 6H polytype. Table \ref{tab:matrix_elements} lists the optical matrix elements and transition energies at $\Gamma$. The alloy direct gap energies increase only slightly from 2H to 6H, reflecting competing effects: reduced hexagonality opens the gap while lower Si content closes it. The interband energies for other transitions decrease from 2H to 6H because the second valence and conduction bands of each polytype fold in from $c_1$ and $v_1$ of more hexagonal polytypes.

\begin{table}
    \centering
    \begin{tabularx}{0.62\textwidth}{XXXXXXX}
    \hline
    \hline
    & \multicolumn{3}{c}{Ge} & \multicolumn{3}{c}{Si$_x$Ge$_{1-x}$}\\
    \cmidrule(lr){2-4}
    \cmidrule(lr){5-7}
    & \multicolumn{1}{c}{2H} & \multicolumn{1}{c}{4H} & \multicolumn{1}{c}{6H} & \multicolumn{1}{c}{2H} & \multicolumn{1}{c}{4H} & \multicolumn{1}{c}{6H}\\
    \midrule
    $E_{11}$ & 0.29 & 0.47 & 0.52 & 0.57 & 0.60 & 0.60\\
    $p_{11\parallel}$ & 0.00 & 0.00 & 0.00 & $3\times 10^{-5}$ & $5\times 10^{-5}$ & $8\times 10^{-5}$\\
    $p_{11\perp}$ & $6\times 10^{-3}$ & 0.00 & $3 \times 10^{-3}$ & 0.173 & 0.133 & 0.124\\
    $E_{21}$ & 0.59 & 0.62 & 0.58 & 1.45 & 0.97 & 0.78\\
    $p_{21\parallel}$ & 0.00 & 0.00 & 0.00 & $9\times 10^{-5}$ & $1\times 10^{-4}$ & $1\times 10^{-4}$\\
    $p_{21\perp}$ & 0.40 & 0.42 & 0.41 & 0.33 & 0.33 & 0.17\\
    $E_{12}$ & 0.41 & 0.55 & 0.58 & 0.70 & 0.70 & 0.68\\
    $p_{12\parallel}$ & 0.00 & 0.00 & 0.00 & 0.08 & 0.11 & 0.12\\
    $p_{12\perp}$ & 0.00 & 0.00 & 0.00 & 0.15 & 0.11 & 0.09\\
    $E_{22}$ & 0.71 & 0.70 & 0.64 & 1.58 & 1.06 & 0.85\\
    $p_{22\parallel}$ & 0.34 & 0.43 & 0.44 & 0.24 & 0.28 & 0.16\\
    $p_{22\perp}$ & 0.35 & 0.31 & 0.29 & 0.31 & 0.27 & 0.13\\
    \hline
    \hline
    \end{tabularx}
    \caption{Interband energies $E_{mn}$ (in eV) and corresponding optical transition matrix elements $p_{\parallel/\perp}$ ($v_n \rightarrow c_m$ in $\hbar/a_B$) at the $\Gamma$-point for pure and alloyed Ge polytypes, showing light polarization perpendicular and parallel to the hexagonal c-axis. Forbidden transitions are denoted by $0.00$. Alloy gap energies represent averages over unique structures for each polytype.}
    \label{tab:matrix_elements}
\end{table}

\subsection{Radiative Lifetime}

The radiative lifetime $\tau$ provides a global measure of optical activity. We computed it by thermally averaging the recombination rate following Ref. \cite{2HClaudia}:
\begin{equation}
    \frac{1}{\tau} = \sum_{c_mv_n\mathbf{k}} A_{mn\mathbf{k}} \frac{w_{\mathbf{k}}e^{-(\varepsilon_m(\mathbf{k})-\varepsilon_n(\mathbf{k}))/(k_{\mathrm{B}}T)}}{\sum_{c_{m'}v_{n'}\mathbf{k}'}w_{\mathbf{k}'}e^{-(\varepsilon_{m'}(\mathbf{k}')-\varepsilon_{n'}(\mathbf{k}'))/(k_{\mathrm{B}}T)}}\text{,}
\end{equation}
where $w_{\mathbf{k}}$ are the $\mathbf{k}$-point weights, $\varepsilon$ the band energies and $A_{mn\mathbf{k}}$ the recombination rates
\begin{equation}
    A_{mn\mathbf{k}} = n_{\mathrm{eff}}\frac{e^2(\varepsilon_m(\mathbf{k})-\varepsilon_n(\mathbf{k}))}{\pi\varepsilon_0\hbar^2m^2c^3} \frac{1}{3} \sum_{i=x,y,z}\left|\left<c_m,\mathbf{k}\left|p_i\right|v_n,\mathbf{k}\right>\right|^2
\end{equation}
with the refractive index $n_{\mathrm{eff}}$ of the surrounding medium (set to 1).

\begin{figure}
    \centering
    \includegraphics[width=0.5\linewidth]{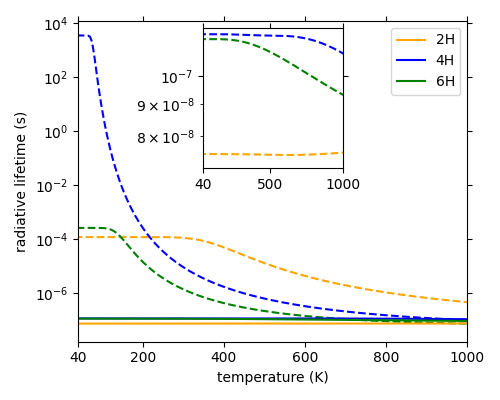}
    \caption{Radiative lifetimes of pure (dashed lines) and alloyed (solid lines) Ge polytypes as functions of temperature.}
    \label{fig:lifetime}
\end{figure}

Figure \ref{fig:lifetime} shows the temperature dependence of radiative lifetimes. Pure hexagonal Ge exhibits the characteristic long lifetime of approximately \SI{0.1}{ms} at low temperatures \cite{2HClaudia}, decaying as temperature increases. While this value significantly exceeds lifetimes measured in real 2H-Ge core-shell nanowires \cite{nature}, other calculations confirm it \cite{Broderick2024arxiv}. The 6H-Ge lifetime behaves similarly but begins decaying at lower temperatures. In strong contrast, the low-temperature lifetime of 4H-Ge is seven orders of magnitude longer, as a direct consequence of the parity-forbidden fundamental transition. This renders radiative recombination virtually impossible in pure 4H-Ge.

Si substitution dramatically alters this picture. Replacing one Ge atom with Si reduces radiative lifetimes by approximately three orders of magnitude (or ten orders for 4H) at 0 K, yielding values of \SIrange{74}{117}{\nano\second}. For 2H alloys, the lifetime remains nearly temperature-independent, while 4H and 6H alloys show temperature dependence similar to pure Ge. The differences between polytypes largely vanish, though the low-temperature ordering persists. These lifetime ratios for pure and alloyed 2H- and 4H-Ge agree qualitatively with experimental measurements \cite{nature}. The dramatic low-temperature lifetime differences between 4H-Ge and the 2H or 6H polytypes shown in Fig. \ref{fig:lifetime} align qualitatively with photoluminescence measurements reporting $\tau = \SI{0.21}{\mathrm{ms}}$ for 4H-Ge \cite{thesis:victor}.

\begin{figure*}
    \centering
    \includegraphics[width=\textwidth]{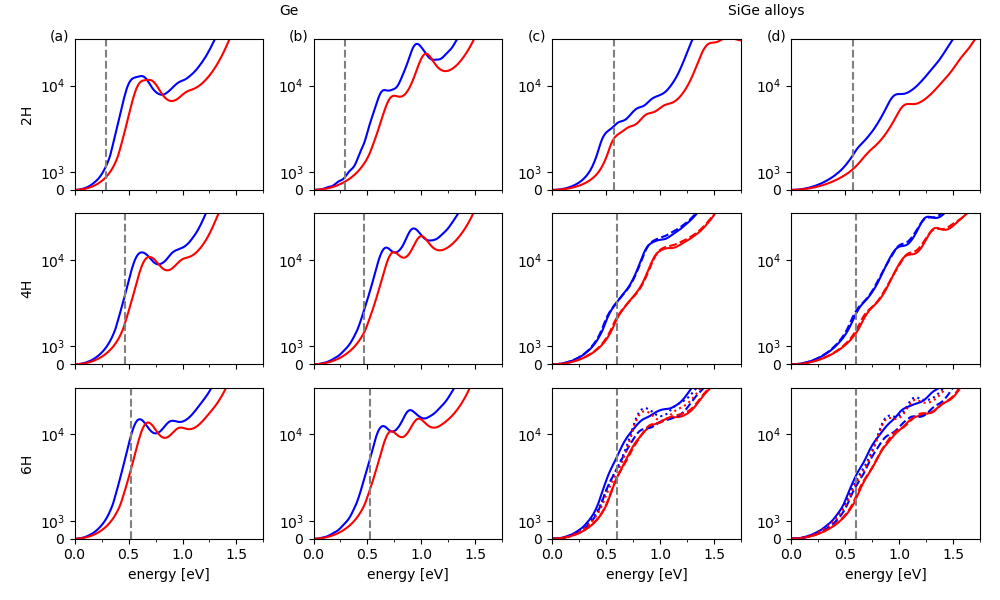}
    \caption{Absorption onset near the fundamental gap (grey dashed line) in cm$^{-1}$ for the 2H, 4H, and 6H polytypes of Ge and SiGe alloys. Red lines show results within the approximate quasiparticle MBJLDA method; blue lines include excitonic effects via the Bethe-Salpeter equation. Panels (a, c) show light polarized perpendicular to the c-axis; (b, d) show parallel polarization. Dashed and dotted blue lines distinguish different Si substitution sites in 4H and 6H systems.}
    \label{fig:absorption}
\end{figure*}

\subsection{Absorption Edge}

We investigated the absorption edges of hexagonal Ge and SiGe materials by solving the Bethe-Salpeter equation to capture excitonic effects mediated by electron-hole interactions. Figure \ref{fig:absorption} compares results including excitonic effects (blue lines) with the independent-quasiparticle MBJLDA approximation (red lines) for all three polytypes. Excitonic interactions generally produce a small redshift of approximately \SI{0.1}{eV} in all spectra. This effect appears most pronounced in the first absorption peak of pure Ge polytypes, which primarily originates from optical transitions between the two highest valence bands and the second-lowest conduction band at energies $E_{21}$ and $E_{22}$ (though $E_{21}$ contributes only to perpendicular polarization). Remarkably, the absorption coefficients $\alpha_{\perp/\parallel}(\omega)$ reach $10^4\,\mathrm{cm}^{-1}$ above photon energies $\hbar\omega = \SI{0.5}{\mathrm{eV}}$ for most polytypes and Si compositions, with values comparable to 3C-GaAs above \SI{1.5}{eV} \cite{Demtroeder3}. Only 2H- and 4H-SiGe spectra require higher photon energies ($\hbar\omega\approx\SI{1}{\mathrm{eV}}$) to reach these values.

The alloy polytypes exhibit a similar peak at their respective transition energies $E_{21}$ and $E_{22}$. However, while the fundamental transition at $E_{11}$ and the next-highest transition at $E_{12}$ remain invisible in pure Ge spectra, consistent with their vanishing or near-vanishing oscillator strengths (Table \ref{tab:matrix_elements}), they appear in the alloys. These transitions emerge as small features for perpendicular polarization (and, for $E_{12}$, parallel polarization as well), matching the oscillator strength behavior. A second, smaller peak appears near \SI{1}{eV} in all pure Ge spectra, corresponding to transitions between the two lowest conduction bands and the third-highest valence band. This feature persists as a shoulder in the alloy systems, though in 2H and 4H alloys it involves only $c_1$. The 6H alloy shows multiple closely-spaced valence and conduction bands that nearly form a plateau in the absorption spectrum.

We observe no excitonic bound states associated with direct gap transitions below the quasiparticle gaps. The visible absorption tails arise primarily from the \SI{0.1}{eV} lifetime broadening applied in our calculations. These excitonic effects, essentially limited to the observed redshift, agree with previous theoretical studies \cite{absorptionEELS,alloyVibrationTheory}.

\section{Dielectric Function and Optical Spectra across the Ultraviolet Range}

\subsection{Imaginary Part and Joint Density of States}

\begin{figure}
    \centering
    \includegraphics[width=\textwidth]{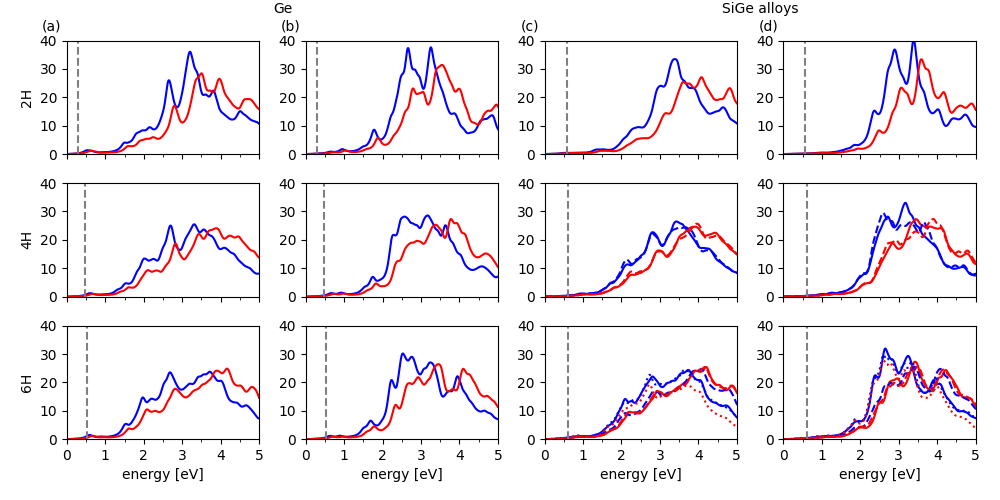}
    \caption{Imaginary part of the dielectric function computed within the approximate quasiparticle MBJLDA method (red lines) and including excitonic effects (blue lines) for the 2H, 4H, and 6H polytypes of Ge and SiGe alloys. Panels (a, c) show light polarized perpendicular to the hexagonal $c$-axis; (b, d) show parallel polarization. Dashed and dotted blue lines distinguish different Si substitution sites in 4H and 6H systems. Grey dashed lines mark the fundamental gap.}
    \label{fig:diel}
\end{figure}

We calculated  the optical properties of nH-Ge and nH-Si$_x$Ge$_{1-x}$ ($n=2,4,6$) through the two diagonal components $\epsilon_{\perp/\parallel}(\omega)$ of the dielectric tensor across a wide energy range extending to the vacuum-UV threshold. We present spectra computed both within the independent-quasiparticle approximation and with excitonic effects included via the Bethe-Salpeter equation. 

The imaginary parts of the dielectric functions shown in Fig.~\ref{fig:diel} are governed primarily by the joint density of states (JDOS), with modulations from the optical transition matrix elements [Eq.~(\ref{eq:momentum_matrix})]. All spectra exhibit the characteristic excitonic redshift caused by attractive electron-hole interactions, accompanied by minor spectral redistribution that enhances peak intensities. This behavior indicates convergence also for 6H, despite the 120 included 3$d$ bands not contributing significantly to the displayed spectral range. In 6H-SiGe, the reduced number of $d$ bands allows more $s$- and $p$-derived valence and conduction bands to contribute to $\mathrm{Im}\,\epsilon_{\perp/\parallel}(\omega)$ above the fundamental gap. Light polarization relative to the $c$-axis significantly affects individual spectral features but has less impact on the overall spectral shape.

The most prominent feature is a double peak in the \SIrange{2.5}{3.5}{eV} range (with excitonic effects included), redshifted by \SIrange{0.1}{0.3}{eV} relative to the independent-quasiparticle spectra of pure Ge polytypes. Si substitution produces an opposite blueshift of approximately \SI{0.5}{eV} in 2H-SiGe alloys, though this shift nearly vanishes for the lower Si concentrations in 4H and 6H. These spectral features arise primarily from van Hove singularities in the JDOS \cite{Demtroeder3}. The double-peak structure resembles van Hove singularities measured by spectroscopic ellipsometry in 3C-Ge \cite{PhysRevB.30.1979,Fernando2017AppliedSurfaceScience,Emminger2020TemperatureDependent}, which appear as spin-orbit-split peaks at $E_1=\SI{2.3}{eV}$ and $E_{1}+\Delta_1=\SI{2.5}{eV}$, plus a shoulder at $E_0'=\SIrange{3.2}{3.3}{eV}$.

For 2H-Ge, the observed peak structures likely originate from transitions between the highest valence bands and lowest conduction bands at the A point (approximately \SI{2.2}{eV}) or L point (approximately \SI{3.4}{eV}), as well as $\Gamma$-point transitions from uppermost valence bands to the third/fourth conduction bands (approximately \SI{2.2}{eV}) or from the third $\Gamma_{7-}^{+}$ valence band to the fifth conduction band (approximately \SI{3.3}{eV}, see the band structure in Fig.~3(b) of Ref.~\cite{2HClaudia}). These transitions produce M$_0$-type van Hove singularities. Our calculated 2H-Ge spectra (Fig.~\ref{fig:diel}) agree qualitatively with other theoretical results for both polarizations \cite{alloyVibrationTheory}, though our peak intensities are somewhat lower than those computed by Bao et al.~\cite{alloyVibrationTheory} while matching the maximum values measured for 3C-Ge \cite{Emminger2020TemperatureDependent}. Notably, the pronounced $E_2$ peak at \SIrange{4.5}{4.6}{eV} observed experimentally in 3C-Ge \cite{PhysRevB.30.1979,Fernando2017AppliedSurfaceScience,Emminger2020TemperatureDependent} is absent from our 2H-Ge theoretical spectra, suggesting a substantial change in optical properties between the cubic and hexagonal phases in this energy region.

\subsection{Absorption Coefficients}

\begin{figure}
    \centering
    \includegraphics[width=\textwidth]{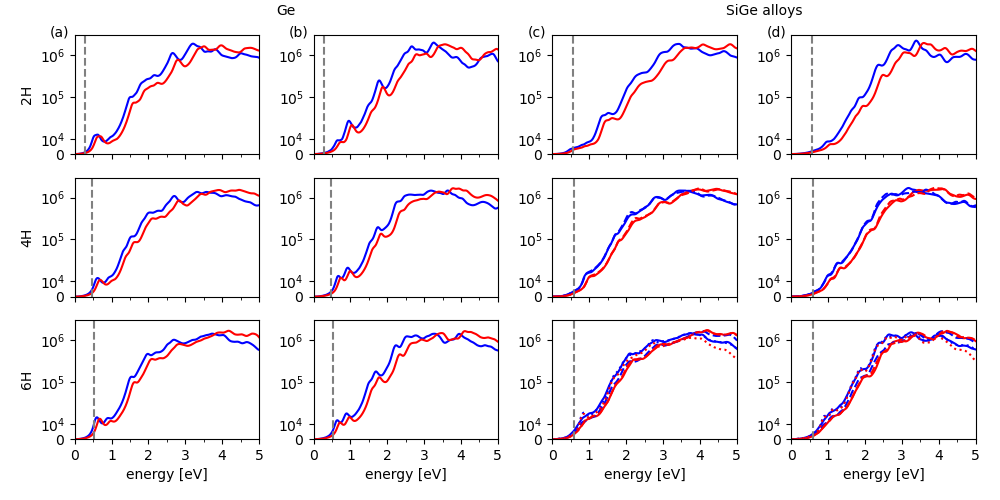}
    \caption{Absorption coefficients (in cm$^{-1}$) computed within the approximate quasiparticle MBJLDA method (red lines) and including excitonic effects (blue lines) for the 2H, 4H, and 6H polytypes of Ge and SiGe alloys. Panels (a, c) show light polarized perpendicular to the $c$-axis; (b, d) show parallel polarization. Dashed and dotted blue lines distinguish different Si substitution sites in 4H and 6H systems. Grey dashed lines mark the fundamental gap.}
    \label{fig:absorption_wide}
\end{figure}

The imaginary parts of the dielectric function (Fig.~\ref{fig:diel}) dominate the absorption coefficients $\alpha_{\perp/\parallel}(\omega)$ displayed in Fig.~\ref{fig:absorption_wide} across most of the spectral range. Above photon energies of approximately $\hbar\omega = \SI{2.5}{eV}$, however, variations in the real parts of the dielectric functions (Fig.~\ref{fig:real}) begin to significantly modify the absorption behavior. In these higher energy ranges, following the strong increase above onset (discussed in Fig.~\ref{fig:absorption}), the absorption coefficients plateau near $\alpha_{\perp/\parallel}(\omega)\approx 10^6\,\mathrm{cm}^{-1}$, largely independent of polytype and Si content. For comparison, experimental absorption coefficients of 3C-Ge reach $10^5\,\mathrm{cm}^{-1}$ at $\hbar\omega = \SI{2}{eV}$, while 3C-Si values are approximately one order of magnitude lower \cite{PhysRev.99.1151}.

\subsection{Real Part and Static Dielectric Properties}

\begin{figure}
    \centering
    \includegraphics[width=\textwidth]{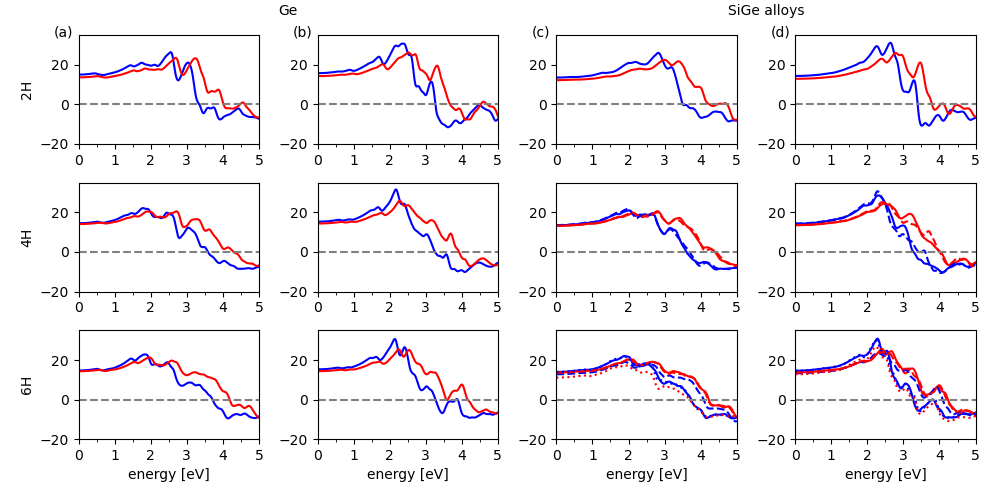}
    \caption{Real part of the dielectric function computed within the approximate quasiparticle MBJLDA method (red lines) and including excitonic effects (blue lines) for the 2H, 4H, and 6H polytypes of Ge and SiGe alloys. Panels (a, c) show light polarized perpendicular to the $c$-axis; (b, d) show parallel polarization. Dashed and dotted blue lines distinguish different Si substitution sites in 4H and 6H systems. The grey dashed line marks $\mathrm{Re}\,\epsilon_{\perp/\parallel}(\omega)=0$.}
    \label{fig:real}
\end{figure}

The real parts of the dielectric functions (Fig.~\ref{fig:real}) influence not only the absorption coefficients but also determine fundamental material properties. Since chemical bonding in the hexagonal polytypes and Ge-rich alloys is essentially covalent, the zero-frequency limit $\lim\limits_{\omega\to 0}\mathrm{Re}\,\epsilon_{\perp/\parallel}(\omega)=\epsilon_{\perp/\parallel\,0}$ yields not only the static electronic dielectric constants, but also the static dielectric constant in general. Table~\ref{tab:dielectric} lists these values computed in both many-body approximations, with and without excitonic effects. Especially for perpendicular light polarization our calculated values fall somewhat below the experimental value $\epsilon_0 = 15.8$ for 3C-Ge \cite{Demtroeder3}, likely reflecting incomplete convergence with respect to the number of bands included. High-energy oscillators needed to satisfy the oscillator-strength sum rule \cite{manybodyElectronicExcitations} remain absent from our finite basis. Nevertheless, the polarization anisotropy should be accurately captured. We consistently find $\epsilon_{\perp\,0}<\epsilon_{\parallel\,0}$ regardless of Si composition, polytype, or the inclusion of excitonic effects, confirming positive uniaxial anisotropy in these materials.

Table~\ref{tab:dielectric} also lists the photon energies $\hbar\omega_{\perp/\parallel\,0}$ at which $\mathrm{Re}\,\epsilon_{\perp/\parallel}(\omega)$ vanishes. Within the simplified Penn model \cite{Ravindra1979}, $\hbar\omega_0$ represents an averaged gap of the semiconductor. Our values lie somewhat below the Penn gaps of $\hbar\omega_0 = \SI{4.3}{eV}$ for 3C-Ge and \SI{4.8}{eV} for 3C-Si \cite{Demtroeder3}. Consequently, applying the Penn model [Eq.~(1) in Ref.~\cite{Ravindra1979}] would yield plasma frequencies below the reference value of \SI{15.6}{eV} for 3C-Ge \cite{Ravindra1979}.

\begin{table}
    \centering
    \begin{tabular}{c c c c c c}
    \hline
    \hline
    material & polytype & \multicolumn{2}{c}{dielectric constant} & \multicolumn{2}{c}{energy zero (eV)}\\
    & & $\perp$ & $\parallel$ & $\perp$ & $\parallel$\\
    \hline
    Ge & 2H & 15.0(13.6) & 15.7(14.2) & 3.34(4.02) & 3.27(3.68)\\
    & 4H & 14.5(14.1) & 15.3(14.4) & 3.58(4.29) & 3.26(3.93)\\
    & 6H & 14.7(14.3) & 15.3(14.4) & 3.71(4.19) & 3.27(3.57)\\
    Si & 2H & 13.5(12.2) & 14.3(12.9) & 3.51(4.24) & 3.39(3.79)\\
    & 4H & 13.5(13.2) & 14.3(13.6) & 3.65(4.42) & 3.35(4.01)\\
    & 6H & 13.6(12.8) & 14.1(13.4) & 3.85(4.04) & 3.56(3.89)\\
    \hline
    \hline
    \end{tabular}
    \caption{Static dielectric constants $\epsilon_{\perp/\parallel\,0}$ and photon energies $\hbar\omega_{\perp/\parallel\,0}$ (in eV) where $\mathrm{Re}\,\epsilon_{\perp/\parallel}(\omega)=0$, extracted from Fig.~\ref{fig:real}. Values with (without) excitonic effects are shown with (in parentheses). Alloy values represent arithmetic averages over unique configurations. For 6H-Ge and 6H-SiGe, only the first zero crossing is listed.}
    \label{tab:dielectric}
\end{table}

\section{Summary and Conclusions}

We have revealed a fundamental difference in the optical properties of hexagonal germanium polytypes that has critical implications for their use in silicon-compatible optoelectronics. Through comprehensive first-principles calculations combining approximate quasiparticle methods with solutions of the Bethe-Salpeter equation, we characterized the optical properties of 2H-, 4H-, and 6H-Ge and their weakly Si-alloyed counterparts across energy ranges from the absorption edge through the ultraviolet.

Our central finding is that 4H-Ge possesses a parity-forbidden fundamental transition. This selection rule that renders its radiative lifetime seven orders of magnitude longer than in 2H- and 6H-Ge. This parity matching between the valence band maximum and conduction band minimum at $\Gamma$ effectively eliminates light emission in pure 4H-Ge, disqualifying it from active optoelectronic applications despite its direct band gap. This discovery explains the experimental observation of microsecond-scale radiative lifetimes in 4H-Ge nanowires, contrasting sharply with nanosecond lifetimes in 2H-Ge.

Critically, we demonstrate that this fundamental limitation can be overcome through controlled symmetry breaking. Single-atom Si substitutions per unit cell break the inversion symmetry, lifting the parity-forbidden character and enhancing optical matrix elements by up to two orders of magnitude. The resulting alloys---2H-Si$_{0.25}$Ge$_{0.75}$, 4H-Si$_{0.125}$Ge$_{0.875}$, and 6H-Si$_{0.083}$Ge$_{0.917}$---exhibit radiative lifetimes in the \SIrange{74}{117}{\nano\second} range, rivaling direct-gap III-V semiconductors while maintaining compatibility with silicon technology. This dramatic transformation from optically inactive to active material establishes alloying as an essential design principle for hexagonal germanium optoelectronics.

Beyond the fundamental transition, we provide comprehensive optical characterization across the full spectral range relevant to device applications. Our calculated absorption coefficients reach $10^4\,\mathrm{cm}^{-1}$ above \SI{0.5}{eV} for most systems, comparable to GaAs, enabling efficient light absorption in thin-film geometries. The dielectric functions exhibit pronounced double-peak structures at \SIrange{2.5}{3.5}{eV}, arising from van Hove singularities in the joint density of states, with clear excitonic redshifts of \SIrange{0.1}{0.3}{eV}. We find consistent positive uniaxial anisotropy ($\epsilon_{\perp\,0}<\epsilon_{\parallel\,0}$) in all systems, offering opportunities for polarization-sensitive device engineering.

For 6H-Ge, we demonstrate optical activity comparable to 2H-Ge, establishing it as a viable alternative for optoelectronic applications and significantly expanding the materials space for hexagonal group-IV heterostructure design. The systematic variation of band gaps and optical properties across the polytype series, combined with their mutual lattice compatibility, enables sophisticated band engineering in multilayer structures.

These results establish hexagonal germanium alloys as a versatile platform for silicon-integrated photonics, where composition and polytype selection offer independent control over emission wavelength and optical strength. Our identification of symmetry-breaking as the key to activating optical transitions in otherwise forbidden systems provides a general design principle applicable beyond the specific materials studied here. Future work should further explore strain engineering, which offers an alternative route to symmetry breaking, and investigate the interplay between alloy disorder and excitonic effects in quantum-confined heterostructures.

\begin{acknowledgments}
We acknowledge financial support from the EU through H2020-FETOpen project OptoSilicon (Grant Agreement No. 964191).
\end{acknowledgments}

\bibliography{main}

\end{document}